%% file: TSMC.tex
\documentclass[lettersize,journal]{IEEEtran}
\IEEEoverridecommandlockouts
\usepackage{cite}
\usepackage{amsmath,amssymb,amsfonts}
\usepackage{amsthm}
\usepackage{algorithm} 
\usepackage{algpseudocode} 
\usepackage{graphicx}
\usepackage{textcomp}
\usepackage{xcolor}
\usepackage{tcolorbox}
\usepackage{tikz}
\usepackage[hyphens,spaces,obeyspaces]{url}
\usepackage[utf8]{inputenc}
\newtheorem{theorem}{Theorem}[section]

\newtheorem{definition}{Definition}[section]

\newtheorem{lemma}[theorem]{Lemma}

\usetikzlibrary{shapes,arrows}
\usetikzlibrary{arrows.meta,calc,fit,positioning}

\tikzstyle{block} = [square, draw,
    text width=7em, text centered, minimum height=4em]
\tikzstyle{sum} = [draw, circle, node distance=3cm]
\tikzstyle{input} = [coordinate]
\tikzstyle{output} = [coordinate]
\tikzstyle{pinstyle} = [pin edge={to-,thin,black}]
\tikzstyle{line} = [draw, -latex']

\tikzset{
   bigbigbox/.style = {minimum width=2.5cm, rectangle},
   bigbox/.style = {draw,rectangle},
   box/.style = {minimum width=2.7cm, rounded corners,rectangle, fill=blue!20},
   square/.style = {minimum width=15mm,minimum height=15mm}
   }

\def\BibTeX{{\rm B\kern-.05em{\sc i\kern-.025em b}\kern-.08em
    T\kern-.1667em\lower.7ex\hbox{E}\kern-.125emX}}
\begin{document}
\bibliographystyle{IEEEtran}
\title{A ZK-SNARK based Proof of Assets Protocol for
Bitcoin Exchanges

}

\author{B Swaroopa Reddy \\ee17resch11004@iith.ac.in}
\markboth{}%
{swaroopa \MakeLowercase{\textit{et al.}}: A ZK-SNARK based Proof of Assets Protocol for
Bitcoin Exchanges}

\maketitle

\begin{abstract}
This paper proposes a protocol for Proof of Assets of a bitcoin exchange using the Zero-Knowledge Succinct Non-Interactive Argument of Knowledge (ZK-SNARK) without
revealing either the bitcoin addresses of the exchange or balances associated with those addresses. The proof of assets is a mechanism to prove the total value of bitcoins the exchange
has authority to spend using its private keys. We construct a privacy-preserving ZK-SNARK proof system to prove the knowledge of the private keys corresponding to the bitcoin assets
of an exchange. The ZK-SNARK tool-chain helps to convert an NP-Statement for proving the knowledge of the private keys (known to the exchange) into a circuit satisfiability problem. In this protocol, the exchange creates a Pedersen commitment to the value of bitcoins associated with each address without revealing the balance. The simulation results show that the proof generation time, size, and verification time are efficient in practice.
\end{abstract}


\section{Introduction}
Blockchain technology gained popularity due to its immutability, trustlessness, and decentralized architecture. Every public blockchain network is associated with a corresponding virtual currency named cryptocurrency (or shortly crypto). Satoshi Nakamoto introduced the first crypto called bitcoin
with the deployment of the bitcoin blockchain \cite{bitcoin} in 2009. The blockchain networks issue  cryptocurrency through a mechanism known as the mining process, e.g., the Proof-of-Work \cite{bitcoin} mechanism in bitcoin. The field of cryptocurrencies is ever-expanding, and as of today, there are more than $4000$ cryptocurrencies in existence. Bitcoin has achieved one trillion-dollar market capitalization \cite{coincap} as there is huge demand from institutional and retail investors. Other popular cryptocurrencies are Ethereum \cite{ETH}, Ripple \cite{XRP}, Zerocash \cite{ZEC}, Stellar \cite{XLM}, Monero \cite{XMR}, etc. 

In a blockchain network, every owner of the crypto holds
a private key to spend the crypto through a chain of digital signatures \cite{bitcoin}. If private keys are stolen or misplaced or the device where the private key stored crashes, the owner loses crypto ownership. So, the users prefer to keep their crypto holdings with exchanges like coinbase \cite{coinbase}, binance \cite{Binance}, etc. The crypto exchanges facilitate crypto trading for fiat currencies or other cryptocurrencies and gain profits through commissions/brokerage charges, listing charges, etc. The exchanges act as an intermediary between buyer and seller by using the mechanism of order-book, similar to
the traditional stock exchanges.
 
The crypto exchanges accept deposits from users through bank transfers or other standard means of deposit. The exchanges hold the private keys on behalf of the users and provides authentication facility through username and password to authenticate the customer’s identity and also provide the recovery facility in case of customer forgets or lost authentication details. So, the customers are free from storing private keys for their cryptocurrencies. But there is a risk of missing customer assets maintained by exchanges as in the case of Mt.Gox exchange \cite{MtGox} due to internal or external frauds. In the traditional banking system, the \textit{central bank} imposes restrictions on the \textit{commercial banks} to maintain a fraction of their total liabilities called fractional reserve ratio \cite{FRR} as reserves, expecting that only a fraction of depositors seek to withdraw funds at the same time. But, the crypto community is expecting a fully solvent exchange instead of proving a fractional solvency of exchange's reserves.

In this paper, we propose a proof of assets protocol for a bitcoin exchange based on the ZK-SNARK proof system \cite{Pinocchio}, \cite{Groth16}. ZK-SNARK is an advancement in the zero-knowledge proofs. Zcash protocol proposed in \cite{zcash} uses ZK-SNARK for constructing the decentralized anonymous payments.
ZK-SNARK is a succinct, non-interactive zero-knowledge proof which facilitates the public verifiability of the proof of a witness. ZK-SNARKs enable the prover to convince the verifier on any non-deterministic decision circuits (NP-statements) with auxiliary information (witness) and public inputs without revealing the witness. A trusted third party takes the circuit as input and generates a common reference string (CRS) consisting of proving and verification keys needed to prove and verify the statement of the ZK-SNARK scheme. 

In this framework, we define the non-deterministic circuit as a statement for verifying the knowledge of all the private keys owned by the exchange to prove the value of bitcoin assets held by the exchange.
The exchange acts as a prover of its total assets and the customers of the exchange play the role of a verifier in the ZK-SNARK proof system. The proof of exchange assets is equivalent to proving ownership of the private keys associated with the bitcoin addresses\footnote{In this work, the bitcoin addresses are the P2PK (Pay to Public Key) addresses \cite{P2PK} where the public keys corresponding to the private keys are called the bitcoin addresses.} owned by an exchange to match the liabilities of the exchange to the customers. 

The exchange as a prover takes the private keys as an auxiliary input, public keys and the corresponding balances as public inputs. The exchange take the proving key and the inputs (auxiliary and public) as input parameters and constructs the proof for the witness. The exchange outputs a Pedersen commitment \cite{Pedersen} to the balance associated with the key pair (part of the auxiliary input). The customers are the public verifiers for verifying the knowledge of the private key to acknowledge the reserves/assets of the exchange from the proof generated by exchanges. The proof size is succinct and it does not leaks the private keys (witness) used in the proof construction. 
The proposed protocol also preserves the privacy of the exchange as the proof system neither reveals the bitcoin addresses information nor the value of the bitcoins held by the exchange.

The results demonstrate that the construction of the proof requires a few hours on an ordinary computer which could be reduced further on a server with high-end processors, which allows the prover to generate proof of assets very frequently. The ZK-SNARK system generates the proof of size, approximately 128 bytes per private key, and the customer can verify the proof of exchange assets in the order of minutes. 

The rest of the paper is organized as follows - In section II, we discuss the related work. Section III describes the preliminaries. In section IV, we discuss the proposed Proof of Assets protocol. In section V, we present the results and discussion. Section VI concludes the paper and gives future directions of the research.

\section{Related Work}
In \cite{maxwell}, the authors discuss the proof of reserves for bitcoin exchange. The maxwell’s proof of reserves discloses the number of bitcoins an exchange holds and the bitcoin addresses for which it knows the private keys. This framework uses the Merkle tree approach to prove the exchange's liabilities by including each
customer’s funds as a leaf of the Merkle tree. The proof of assets is a straightforward approach of providing signatures with all private keys owned by the exchange.

In Provisions \cite{Provisions}, the authors propose a privacy-preserving proof of solvency for bitcoin exchanges using crypto primitives like zero-knowledge proofs \cite{Fiat-Shamir, Schnorr}, and Pedersen
commitments \cite{Pedersen}. Provisions discusses three protocols – Proof of assets, proof of liabilities, and proof of solvency. It also
discusses the proof of non-collusion between exchanges.
The proof of assets $\Sigma$ protocol proves the knowledge of exchange's assets by providing Pedersen commitments to the amounts of bitcoins the exchange holds for a set of known public keys. It also proves the knowledge of a binary value using Pedersen commitments if it knows the private keys corresponding to the known public keys. The proof of liabilities provides the Pedersen commitments to balances of each customer associated with the exchange and the Pedersen commitments to the bits of the binary representation of the customer’s balance.
The proof of solvency proves that the difference
between the assets and liabilities is either a zero or a positive (if an exchange is a surplus). In the optional proof of non-collusion protocol, the exchange creates two lists to prove the knowledge of a non-collision with any other exchange. The first list consists of Pedersen commitments to the private keys corresponding
to the known public keys and the second list consists of public keys generated with a base change. The exchange proves that the second list is a permutation, unblinding, and base change of the first list.

In \cite{blockstream}, an exchange constructs a transaction as proof of reserves with all the bitcoin UTXOs spendable by the exchange by adding an extra invalid input such that the exchange is unable to spend its own UTXOs.  So, this approach discloses all the UTXOs of the exchange along with the public keys
owned by an exchange.

\section{Preliminaries}
\label{prelim}
In this section, we describe the background on cryptographic primitives used in the protocol - Ellicptic curves, Pedersen commitments and ZK-SNARK. In this work, we stick to the Elliptic curve cryptography \cite{ecc} used in bitcoin to prove the ownership of private keys corresponding to the bitcoin addresses.

\subsection{Point Addition and Point Double}
Let $\mathbb{E}$ (including a point at infinity $\mathcal{O}$) be a group of order $q$ corresponding to points on the elliptic curve $y^2 = x^3 + ax + b$ over a finite field $\mathbb{F}_p$. The addition operation $+$ on $\mathbb{E}$ \cite{ecc-add}  is defined as follows - 

If $P(x_1,y_1)$ and $Q(x_2,y_2)$, then $P+Q = (x_3, y_3)$, where
\begin{equation}
x_3 = m^2 - (x_1 + x_2),  \hspace{0.2cm} y_3 = m(x_1 - x_3) - y_1
\label{eq:ecc_add}
\end{equation}
and,
\[ m = 
\begin{cases}
    \frac{y_2-y_1}{x_2-x_1},& \text{if } P \neq Q\\
    \frac{3x_1^2+a}{2y_1},           & \text{otherwise}
\end{cases}
\]

\subsection{Bitcoin public and private keys}
Let $G \in \mathbb{E}$ be a generator or base point. The public key $K$ (bitcoin address) corresponding to the private key $k \in \mathbb{Z}_q = \{1, 2, \dots, q-1\}$ is calculated as a scalar multiplication \cite{ecc-add} of $k$ with $G$.

\begin{equation}
K = kG = G+G+ \dots + G \hspace{0.1cm }(k \hspace{0.1cm} times)
\end{equation}
Calculating $k$ from $K$ is called the \textit{discrete logarithm problem} \cite{DL} which is assumed to be hard.

\subsection{Pedersen Commitments}
Pedersen commitment \cite{Pedersen} $c \in \mathbb{E}$ is   used to perfectly hide a message $m  \in \mathbb{Z}_q$. Let $H \in \mathbb{E}$ be an another generator independent of $G$ and ensure that the discrete logarithm of $H$ for $G$ is unknown. The Pedersen commitment $c$ to a message $m$ is
\begin{equation}
c = mG + bH
\end{equation}
Where $b \in \mathbb{Z}_q$ is a randomly chosen blinding factor. The commitment $c$ completely hides the message $m$ and it is opened by revealing $m$ and $b$, one can check $ mG + bH \stackrel{?}{=} c$.
The Pedersen commitments are additively homomorphic. 

\subsection{Non-Interactive zero-knowledge argument of knowledge for an NP language \cite{zcash}}
An NP language is a set of statements $L$ such that if an instance $z \in L$, then there exists a witness $w$, that proves the membership of the statement $z$ in $L$ in polynomial time. In a non-interactive zero-knowledge argument of knowledge for an NP language $L$, prover convinces verifier that it has knowledge of auxiliary input $w$ for $z \in L$ without revealing  $w$ to the verifier. In our protocol, we use ZK-SNARK to prove and verify the membership of the instance $z$ in $L$. ZK-SNARK uses a trick to reduce any NP-statement $L$ to  circuit satisfiabilty problem (NP complete problem). The NP-statement is converted to non-deterministic decision circuit $C$ such that the input to the statement is transformed as the input to the circuit $C$.

\subsection{ZK-SNARK}
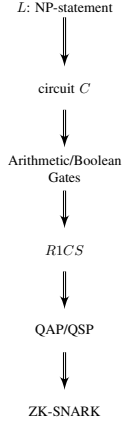
\begin{figure}[t]
\centering
\resizebox{0.25\columnwidth}{!}{
\input{./figs/ZK-SNARK.tex}
}
\caption{Overview of the ZK-SNARK Toolchain}
\label{fig:zk-snark}
\end{figure}

ZK-SNARK \cite{QSP, Pinocchio, Groth16} is a variant of Zero-knowledge proof of knowledge \cite{Fiat-Shamir}, \cite{Schnorr} with succinct proof. ZK-SNARK is used to prove and verify any instance belongs to an NP language $L$.  Let $L$ be an NP statement and $C$ represents the non deterministic decision circuit for the instance $z$. The ZK-SNARK toolchain takes the circuit $C$ of the instance $z$  as input and generates ZK-SNARK proof system as described in Fig. \ref{fig:zk-snark}. 

Initially, the circuit $C$ for $z \in L$ is converted into arithmetic gates consisting of wires with values (from a field $\mathbb{F}_p$) connected to multiplication ($*$) and addition ($+$) gates. Similarly, boolean circuits operate over bits with gates like OR, AND, XOR, etc. Next, we convert the algebraic gates into a Rank1 constraint system (R1CS) \cite{vitalik}, which is a group of three vectors ($\mathbf{a}$, $\mathbf{b}$, $\mathbf{c}$) and $\textbf{t}$ is a solution vector to the R1CS such that
\begin{equation}
\textbf{t}.\textbf{a} * \textbf{t}.\textbf{b} = \textbf{t}.\textbf{c}
\label{eq:constraint}   
\end{equation} 
where $(.)$ represents the dot product. The vector $\textbf{t}$ ensures the satisfaction of the constraint \eqref{eq:constraint}. There are as many constraints as the number of algebraic gates in the system. The length of the vectors is equal to the total number of variables used in the circuit. The detailed explanation of the R1CS constraint system is discussed with examples in \cite{vitalik}.

The next step is converting $R1CS$ into a Quadratic Arithmetic Program (QAP)/ Quadratic Span Program (QSP) \cite{QSP} form to implement the same logic as $R1CS$. A QAP  $Q$ over a field $\mathbb{F}_p$ is defined as a tuple consisting of the three sets of $m+1$ polynomials and a target polynomial
\begin{equation}
Q = \left(\{u_i(x)\}_{i=0}^{m}, \{v_i(x)\}_{i=0}^{m}, \{w_i(x)\}_{i=0}^{m} ,  t(x) \right)
\end{equation}

Suppose, if there exists $l+1$ public inputs and $m-l$ auxiliary inputs (or witness), then the QAP $Q$ computes $C$ iff a prover needs to prove that he knows a secret called witness $w = (a_{l+1}, \dots, a_m) \in \mathbb{F}^{m-l}$, which satisfies the following equation with public inputs $z = (a_0=1,a_1 \dots, a_l) \in \mathbb{F}^{l+1}$ such that $t(x)$ divides $p(x)$ (i.e., $p(x) = t(x)h(x)$), where
\begin{equation}
p(x) = \left(\sum_{i=0}^{m} a_i u_i(x)\right).\left(\sum_{i=0}^{m} a_i v_i(x)\right) - \left(\sum_{i=0}^{m} a_i w_i(x)\right)
\label{eq:QAP}
\end{equation}
The construction of the polynomials of $Q$ is discussed in \cite{QSP, Pinocchio, vitalik}. The polynomials in $Q$ and $h(x)$ are specific to a particular circuit $C$ corresponding to an instance of $L$ and independent of public or auxiliary inputs. Finally, the polynomials of $Q$ are used in the construction of the ZK-SNARK proof system.

More formally, the relation between the circut satisfiability of an arithmentic circuit $C:\mathbb{F}_p^{l+1} \times \mathbb{F}_p^{m-l} \rightarrow \mathbb{F}_p^k$ and NP-language $L$ is defined by the relation $\mathcal{R}:=\{(z,w) \in \mathbb{F}_p^{l+1} \times \mathbb{F}_p^{m-l} \hspace{0.1cm} s.t. \hspace{0.1cm} C(z,w) = \mathbf{0}^k\}$ and its language $L:=\{z \in \mathbb{F}_p^{l+1}: \exists w \in \mathbb{F}_p^{m-l} \hspace{0.1cm} s.t. \hspace{0.1cm} C(z,w) = \mathbf{0}^k\}$.

In this paper, we stick to the ZK-SNARK framework proposed by Groth in \cite{Groth16} as this protocol consists of a shorter proof with $3$ group elements ($2$ elements from group $\mathbb{G}_1$ and $1$ element from  group $\mathbb{G}_2$) compared to $9$ group elements in Pinocchio protocol \cite{Pinocchio}. Also the proof construction and verification times are less in Groth's protocol compared to Pinocchio's protocol. 

ZK-SNARK framework proposed in \cite{Groth16} is a set of three probabilistic polynomial time (PPT) algorithms defined as follows:\\ \\
\textbf{1. Setup Phase: $\sigma \leftarrow Setup(C,1^{\lambda})$}

A trusted third party takes  $C$ as input and generates common reference string $\sigma$ as follows:
\begin{itemize}
\item Pick $\alpha$, $\beta$, $\gamma$, $\delta$, $x \leftarrow \mathbb{F^{*}}$, set $\mathbf{\tau} = (\alpha$, $\beta$, $\gamma$, $\delta$, $x$)
\item compute $\sigma = \left([\sigma_1]_1, [\sigma_2]_2\right)$, where
\begin{align}
\begin{split}
\sigma_1 &= \begin{pmatrix}
			\alpha, \beta, \delta, \left\lbrace x^i\right\rbrace_{i=0}^{n-1}, \left\lbrace \frac{\beta u_i(x) + \alpha v_i(x) + w_i(x)}{\gamma} \right\rbrace_{i=0}^{l} \\
\left\lbrace \frac{\beta u_i(x) + \alpha v_i(x) + w_i(x)}{\delta} \right\rbrace_{i=l+1}^{m}, \left\lbrace \frac{x^i t(x)}{\delta}\right\rbrace_{i=0}^{n-2}
\end{pmatrix}, \\
\sigma_2 &= \begin{pmatrix}
\beta, \gamma, \delta, \left\lbrace x^i\right\rbrace_{i=0}^{n-1}
\end{pmatrix}
\end{split}
\label{eq:CRS}
\end{align}
\end{itemize}
Where, $[.]_1$ denotes elements on the group $\mathbb{G}_1$ and, $[.]_2$ denotes elements on the group $\mathbb{G}_2$. \\ \\
\textbf{2. Proving phase:$(\pi, y) \leftarrow Prove\left(\sigma, \{a\}_{i=0}^{m}\right)$ } 

The prover computes the proof $\pi$ from the public statement $z = \left\lbrace a_i \right\rbrace_{i=0}^{l}$ and witness $w = \left\lbrace a_i \right\rbrace_{i=l+1}^{m}$ as follows
\begin{itemize}
\item Pick randomly $r$, $s \leftarrow \mathbb{Z_p^*}$
\item Compute proof $\pi = \left([A]_1, [B]_2, [C]_1 \right)$, where 
\begin{align}
\begin{split}
A &= \alpha + \sum_{i=0}^{m} a_i u_i(x)  + r\delta , \hspace{0.2cm} B = \beta + \sum_{i=0}^{m} a_i v_i(x)  + s\delta \\ 
C &= \frac{\sum_{i=l+1}^{m} a_i\left(\beta u_i(x) + \alpha v_i(x) + w_i(x)\right)+h(x)t(x)}{\delta} \\
&+ As + Br - rs\delta
\end{split}
\label{eq:pi}
\end{align}
\end{itemize}
This algorithm also computes the output ($y$) of the circuit $C$. \\ \\
\textbf{3. Verification phase: $b \leftarrow Verify\left(\sigma, \{a\}_{i=0}^{l}\right)$, $b \in \{0,1\}$ } 

The verifier use the public inputs $\left\lbrace a_i \right\rbrace_{i=0}^{l}$ to verify the proof $\pi$ and accepts the proof if and only if  
\begin{align}
\begin{split}
[A]_1 . [B]_2 &= [\alpha]_1.[\beta]_2 + [C]_1.[\delta]_2 \\
&+ \sum_{i=0}^{l} a_i\left[\frac{\beta u_i(x) + \alpha v_i(x) + w_i(x)}{\gamma}\right]_1.[\gamma]_2  
\end{split}
\label{eq:verify}
\end{align}
The above construction of the ZK-SNARK protocol is a non-interactive zero-knowledge arguement of knwoledge with perfect completeness and perfect zero-knowledge (Theorem 2 of \cite{Groth16}).

\section{The proposed Proof of Assets Protocol}
The protocol consists of three major entities - Trusted third party, Prover - crypto exchange $\mathcal{E}$ and, Verifier - customer $\mathcal{C}$ of the exchange $\mathcal{E}$, who holds crypto assets with $\mathcal{E}$. In Proof of Assets protocol, the exchange proves the total bitcoins over which it has the ownership authority to spend. In the proposed protocol, the exchange  $\mathcal{E}$ proves its total assets in zero-knoweldge and also it preserves the privacy without revealing its public key addresses and associated balances. 
The exchange $\mathcal{E}$ generates a ZK-SNARK proof for an NP-statement which says $\mathcal{E}$ knows the private keys for a subset of bitcoin addresses (public keys) and also computes a Pedersen commitment to its bitcoin assets.

Let $\mathbf{PK}$ be the set of total bitcoin public keys on the blockchain.
\begin{equation}
\mathbf{PK} = \left\lbrace y_1, y_2, \dots, y_k \right\rbrace  \subseteq \mathbb{E}
\end{equation}
Let $x_1$, $x_2$, $\dots$, $x_k$ $\in \mathbb{Z}_p$ are the set of private keys corresponding to the public keys from the set $\mathbf{PK}$, such that $y_i = x_i G$ for $i=1,2,\dots, k$.

Let $\mathbf{S_{own}}$ be the subset of the public keys for which the exchange knows the private keys and $\mathbf{S_{own}} \subset \mathbf{PK}$. If $\mathcal{E}$ provides proofs only for the private keys associated with addresses in $\mathbf{S_{own}}$, it revals the bitcoin addresses and total assets owned by $\mathcal{E}$. So, $\mathcal{E}$ takes an anonymity set $\mathbf{S_{anon}} = \left\lbrace y_1, y_2, \dots, y_n \right\rbrace$  ($n < k $) such that $\mathbf{S_{anon}} \subset \mathbf{PK}$ and $\mathbf{S_{anon}} \supset \mathbf{S_{own}}$ to prove the assets owned by $\mathcal{E}$. 

Let $s_i \in \left\lbrace 0, 1\right\rbrace$ denotes which public keys the exchange knows the private key. If $s_i = 1$, then the exchange knows the private key $x_i$ corresponding to the bitcoin address $y_i \in \mathbf{S_{own}}$. Let $v_i$ denotes the amount of bitcoins associated with address $y_i \in \mathbf{S_{anon}}$ and $\mathbf{V} = \left\lbrace v_1, v_2, \dots, v_n \right\rbrace$, the total assets of the exchange called $\mathcal{E}_{Assets}$ is defined as

\begin{equation}
\mathcal{E}_{Assets} = \sum_{i=1}^{n} s_iv_i = \sum_{y_i \in \mathbf{S_{own}}}v_i
\end{equation}

\subsection{The NP-statement \textit{PoA} for Proof of Assets}
We construct a Proof of Assets protocol for an exchange to prove the ownership of the bitcoins it hold. We use ZK-SNARK and Pedersen commitment to prove exchange's assets in zero-knowledge.
The ownership of the bitcoin is defined by $s_i$, which can be evaluated by checking the equation $y_i \stackrel{?}{=} x_iG$. To preserve the integity and privacy of the exchange, $x_i$ and  $s_i$ are used as a part of the witness in the construction of the ZK-SNARK proof system. The Pedersen commitment hides $v_i$ through the secret $r_i$. 

Let $F = PoA(z_i, w_i)$ be the NP-statement or the non-deterministic decision function with the public inputs  $z_i = \left\lbrace y_i, v_i \right\rbrace$ and the witness $w_i=\left\lbrace x_i, s_i, r_i\right\rbrace$, then the NP-statement \textit{PoA} for proving the exchange's ownership on $x_i$ is defined as

\textit{"Either I know the private key $x_i$ corresponding to a public key $y_i$ in which case $s_i = 1$ and $c_i$ is the commitment to the value $v_i$ or I don't know the private key $x_i$ corresponding to the public key $y_i$ in which case $s_i = 0$ and $c_i$ is the commitment to the value $0$."}

The statement \textit{PoA} is captured by the relation $\mathcal{R}_{PoA} = \{(z_i,w_i) \in \mathbb{F}_p^{l+1} \times \mathbb{F}_p^{m-l} \hspace{0.1cm} s.t. \hspace{0.1cm} C_{PoA}(z_i,w_i) = c_i\} \}$ and its corresponding language is $L:=\{z_i \in \mathbb{F}_p^{l+1}: \exists w_i \in \mathbb{F}_p^{m-l} s.t. (z_i, w_i) \in \mathcal{R}_{PoA}\}$. Where $C_{PoA}$ is an arithmetic circuit representation of the statement \textit{PoA} and the tuple $(z_i,w_i)$ represents the values assigned to variables $a_j$, for $j = 1,2, \dots, m$ in \eqref{eq:QAP}.

In other words, for a given \textit{PoA} instance $z_i$, the witness $w_i$ is valid iff:
\begin{enumerate}
\item The private key ($x_i$) should match the public key ($y_i$) or not, i.e. $s_i \in \{0,1\}$.
\item The commitment $c_i$ (output of the circuit $C_{PoA}$) is computed correctly, i.e., $c_i \stackrel{?}{=}  s_iv_iG + r_i H$.
\end{enumerate}

\subsection{Arithmentic Circuit $C_{PoA}$ for verifying NP-statement \textit{PoA}}
\label{sec:circuit}
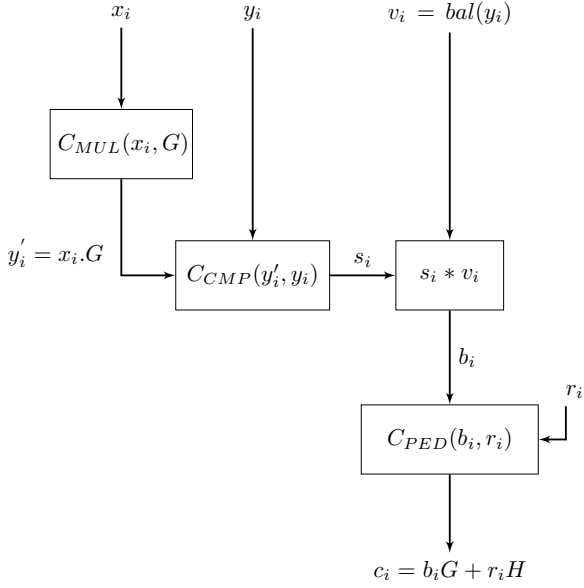
\begin{figure}[t]
\centering
\resizebox{0.9\columnwidth}{!}{
\input{./figs/NP.tex}
}
\caption{Circuit $C_{PoA}$ for the NP-statement $PoA$}
\label{fig:NP-statement}
\end{figure}

As shown in Fig. \ref{fig:zk-snark}, ZK-SNARK is a proof system for arithmetic circuit satisfiability problem to prove the witness $w_i$ for an instance $z_i$. So, we express the checks in the NP-Statement \textit{PoA} to arithmetic circuit \textit{$C_{PoA}$} as  depicted in Fig. \ref{fig:NP-statement}. There are three subcircuits in $C_{PoA}$ - Scalar multiplication circuit $C_{MUL}$, Comparision circuit $C_{CMP}$, and Pedersen commitment circuit $C_{PED}$. Each individual circuit consists of arithmetic gates for $'*'$, $'+'$, $'-'$ or $'/'$.  As per the R1CS constraint equation \eqref{eq:constraint}, each gate can be represented as a tuple $(\textbf{a}, \textbf{b}, \textbf{c},\textbf{t})$ called a constraint.
 
\subsubsection{Scalar Multiplication circuit verification:}
\label{sec:C_MUL}
The circuit $C_{MUL}$ is the scalar multiplication of $x_i$ with base point $G$. It is used to test whether the exchange $\mathcal{E}$ knows the  private key $x_i$ matches to the corresponding public key $y_i$. The scalar multiplication is defined as
\begin{equation}
y_i = x_iG = G + G + \dots + G \hspace{0.2cm} (x_i \hspace{0.1cm} times)
\end{equation}

Where, $x_i$ and $x,y$-coordinates of the $G$ are $256$-bit numbers over the field $\mathbb{F}_p$ used for bitcoin addresses.
Instead of constructing $x_i$ number of point addition circuits of $G$, we construct $C_{MUL}$ from $256$ point addition circuits. Initially, scalar $x_i$ is unpacked into a vector \textit{$xvec_i$} of $256$-bits with $257$ constraints\footnote{Each constraint represnts a single arithmetic gate.}.

Let $P_{add}$ be a point addition circuit. We construct $P_{add}$ gadget with  $3$ arithmetic gates as per point addition  for elliptic curve points defined in \eqref{eq:ecc_add}. We process $2$-bits (out of 256 bits) of $x_i$ at a time to reduce the number of $P_{add}$ gadgets to half of the total number of bits (i.e to $128$) by adding an extra constraint (total of $4$ constraints per $P_{add}$) for the product of the two bits of $x_i$. 

Algorithm \ref{Algorithm1} describes the construction of $C_{MUL}$ circuit using $P_{add}$ circuits. In every iteration, we compute the accumulation of constant $G$ (i.e. $a$, $b$, $c$ for $j=0,\dots, numbits$) as the input to $P_{add}$ circuit. In each iteration, the powers   $a$, $b$, $c$ are computed from their previous values. For simplicity, a point (accumulator) on the curve other than $G$ or $H$ is added to the output of the circuit $C_{MUL}$ and substracted at the end with additional $P_{add}$ circuit. The summary of the total number of constraints (or arithmetic gates) required to construct the proof for  $C_{MUL}$ circuit is listed in Table~\ref{tab1}.  
\begin{table}[!b]
\caption{Size of circuit $C_{MUL}$}
\begin{center}
\begin{tabular}{|l|l|}
\hline
\multicolumn{2}{|c|}{\textbf{Gate count for $C_{MUL}$}} \\
\cline{1-2} 
Unpacking $x_i$ to $xvec_i$ & $257$  \\
Scalar Multiplication ($128$ $P_{add}$ circuits) & $512$  \\
($P_{add}$ (Point addition)) & \hspace{0.15cm}(4) \\
Accumulator ($P_{add}$) & \hspace{0.2cm} $3$ \\
 \hline
\textbf{Total} & $772$ \\
\hline
\end{tabular}
\label{tab1}
\end{center}
\end{table}
\begin{algorithm}[!t]
  \caption{Scalar Mltiplication circuit $C_{MUL}$}\label{Algorithm1}
  \textbf{Auxilary input} - $\left\lbrace x_i\right\rbrace$ \\ 
  \textbf{Public input} - $\left\lbrace numbits, G, y_i\right\rbrace$ 
  \begin{algorithmic}[1]
    \Procedure{$C_{MUL}$}{$x_i$, $G$, $numbits$, $y_i$}
      \State Unpack $x_i$ into a vector $xvec_i$ of length \textit{$numbits$}
      \While{$j < numbits$}
          \State Compute $a = 2^j.1.G$, $b = 2^j.2.G$, $c = 2^j.3.G$     
          \If{$xvec_i[j] == 0 \& xvec_i[j+1] == 0$}
             \State $out[k] = P_{add}(out[k-1], 0)$
          \ElsIf{$xvec_i[j] == 1 \& xvec_i[j+1] == 0$}
             \State $out[k] = P_{add}(out[k-1], a)$
          \ElsIf{$xvec_i[j] == 0 \& xvec_i[j+1] == 1$}
             \State $out[k] = P_{add}(out[k-1], b)$
          \Else
             \State $out[k] = P_{add}(out[k-1], c)$
          \EndIf
          \State $j = j+2$, $k=k+1$       
        \EndWhile
        \State $y_i = out\left[\frac{numbits}{2} - 1\right]$
    \EndProcedure
  \end{algorithmic}
\end{algorithm}

\subsubsection{Comparison circuit verification:}
The comparison function $CMP$ compares the output of the scalar multiplication $y_i^{'}$ with the input $y_i$.
The circuit ensures the comparison of $x$, $y$ coordinates of  $y'_i$ and $y_i$ to yield a binary output $s_i$.
This is achieved using a single constraint to obtain the product of $sx$ and $sy$ as $s$, where $sx$ and $sy$ are the outputs for comparing $x$ and $y$ coordinates of $y'_i$, $y_i$ respectively. We also add a constriant to $C_{CMP}$ circuit to ensures that $s_i \in \{0,1\}$ using the operation $s_i * s_i = s_i$. So, we express the $C_{CMP}$ circuit using two constriants.

\subsubsection{Pedersen Commitment circuit Verification:} $C_{PED}$ circuit evaluates the Pedersen commitment of $b_i = s_i. v_i$ with a randomly chosen blinding factor $r_i$ which yields an output $c_i = b_iG + r_i H$. This circuit requires two $C_{MUL}$ circuits, an arithmetic gate for computing $b_i$. Since the total bitcoin supply is limited to $21$ million coins, the value of $v_i$ is bounded by $2^{z}$, where $z \in [0,51]$. So, the total number of constraints for the first $C_{MUL}$ circuit, i.e., $b_iG$ depends on the maximum value of $z$. Since the blinding factor $r_i \in \mathbb{Z}_q$, the second multiplication is similar to the $C_{MUL}$ circuit discussed in Table~\ref{tab1}.
The summary of the total number of constraints  required to construct a proof for $C_{PED}$ circuit is listed in Table~\ref{tab2}.

\begin{table}[!t]
\caption{Size of circuit $C_{PED}$}
\begin{center}
\begin{tabular}{|l|l|}
\hline
\multicolumn{2}{|c|}{\textbf{Gate count for $C_{PED}$}} \\
\cline{1-2} 
 Arithmetic Gate $\left(b_i = s_i * v_i\right)$& \hspace{0.2cm} $1$ \\
 $C_{MUL}$ for $b_iG$  & $155$  \\
 $C_{MUL}$ for $r_iH$  & $769$  \\
 Accumulator ($P_{add}$) &  \hspace{0.2cm} $3$ \\
\hline
\textbf{Total} & $928$ \\
\hline
\end{tabular}
\label{tab2}
\end{center}
\end{table}
Finally, by combining all the circuits, a total of $1702$ constraints are required to prove circuit \textit{$C_{PoA}$} for NP-statement \textit{PoA}. When $\mathcal{E}$ wants to provide a proof $\pi_i$ for $y_i \in \mathbf{S_{anon}}$ and $y_i \not \in \mathbf{S_{own}}$, it can directly compute the proof with only considering the Pedersen commitment circuit $C_{PED}$ by directly taking $s_i = 0$ to reduce the number of constriants to $928$.
\subsection{The Proof of Assets protocol}
The proof of Assets protocol shown in Table \ref{POA} for bitcoin exchanges is constructed based on the ZK-SNARK framework illustrated in Section \ref{prelim}. The proposed protocol satisfies the following properties.
\begin{itemize}
\item No probabilistic polynomial time (PPT) adversarial exchange can provide a commitment to the balance which exceeds the actual amount it owns as the $v_i$ is a public input to the ZK-SNARK proof system. 
\item No PPT adversarial exchange will be able to provide existence of $v_i$ in $C_{Assets}$ for an address $y_i \not \in \mathbf{S_{own}}$.
\item No PPT adversary will be able to distinguish the address in the $\mathbf{S_{anon}}$ belongs to the $\mathbf{S_{own}}$.

\begin{table}[!t]
\caption{Proof of Assets Protocol}
\begin{tcolorbox}[colback=white!5!white,colframe=white!30!black]
For $i\in[1, n]$ \\
Public input from blockchain: $z_i = (y_i, v_i$)  \\
Exchange's private input (witness): $w_i = (x_i, s_i, r_i)$ \\
Verifier's input from Exchange: $c_i$, $C_{Assets}$
\begin{enumerate}
\item \textbf{Setup phase:} The trusted third party generates the CRS $\sigma$ for the NP-statement \textit{PoA}.
\item \textbf{Proving phase:}
\begin{enumerate}
\item For $i\in[1, n]$ 
\begin{itemize}
\item The exchange $\mathcal{E}$ constructs the proof 
\begin{equation}
\pi_i  \leftarrow Prove(\sigma, z_i, w_i)
\label{eq_proof}
\end{equation}
\item Compute the output from $w_i$ and $z_i$
\begin{align*}
 r_i \stackrel{\$}{\leftarrow} \mathbb{Z}_q
\end{align*}
 \begin{equation}
 c_i = s_iv_iG+r_iH
 \end{equation}  
\end{itemize}
\item Compute the commitment to total assets
\begin{equation}
C_{Assets} = \left(\sum_{i=1}^{n} b_i \right)G + \left(\sum_{i=1}^{n} r_i \right)H
\end{equation}
\end{enumerate}

\item  \textbf{Verification phase:} The customer verify the following
\begin{enumerate}
\item  For $i\in[1, n]$
  \begin{equation}
    b_{1,i} \in \{0,1\} \leftarrow Verify(\sigma, z_i,c_i,\pi_i)
    \label{eq:verify_c_i}
  \end{equation}
  \item \begin{equation}
    b_{2,i} \in \{0,1\}\leftarrow Verify(\sigma, C_{Assets},\pi_i)
    \label{eq:verify_C_Assets}
  \end{equation}
  \item The verifier checks 
\begin{equation}
C_{Assets} \stackrel{?}{=} \sum_{i = 0}^{n} c_i \label{eq:C_Assets}
\end{equation}  
\end{enumerate}
\end{enumerate}
\end{tcolorbox}
\label{POA}
\end{table}

\end{itemize}
\subsubsection{CRS Setup:}
A trusted third party construct a ZK-SNARK's common-reference string $\sigma$
\eqref{eq:CRS} for the NP-statement \textit{PoA} as per the ZK-SNARK toolchain described in Fig. \ref{fig:zk-snark}. First, it construct a QAP $Q$ from the circut $C_{PoA}$ for the statement \textit{PoA}, then it generates $\sigma = \left([\sigma_1]_1, [\sigma_2]_2\right)$ using $Q$ and trapdoor information $\alpha$, $\beta$, $\gamma$, $\delta$, $x \leftarrow \mathbb{Z}_q$. 
\subsubsection{Generation of the proof:}
The crypto exchange $\mathcal{E}$ (Prover) constructs a proof $\pi_i$ to prove the knowledge of a private key $x_i$ corresponding to the public key $y_i$. As per \eqref{eq:pi}, $\mathcal{E}$ needs to provide valid assignments\footnote{Assignments refer to the coeffiecients $a_i$'s that satify the polynomial $P(x)$ in \eqref{eq:QAP}. The problem of polynomial satisfiabilty is converted into proving and verification mechanism shown in \eqref{eq:pi} and \eqref{eq:verify}.} $\{a_{ij}\}_{j=0}^{m}$ to generate $\pi_i = (A_i, B_i, C_i)$, for all $i = [1,n]$. Where, $\{a_{ij}\}_{j=0}^{l}$ represents the public input $z_i = (y_i, v_i)$ from the blockchain and $\{a_{ij}\}_{j=l+1}^{m}$ represents the $\mathcal{E}$'s auxiliary input $w_i=\left\lbrace x_i, s_i, b_i, r_i, t_i\right\rbrace$. Where, $t_i$ is the assignments for all the internal wires of the circuit $C_{PoA}$ excluding $s_i$ and $b_i$. 

The prover also provide output $c_i$ to the verifier, which is a Pedersen commitment to $b_i = s_iv_i$, for all $i = 1, \dots, n$. If the exchange knows the private key $x_i$, then $s_i = 1$ and $c_i$ is the commitment to balance $v_i$, other wise it is a commitment to value zero.

\begin{equation}
c_i= 
\begin{cases}
    v_iG+r_iH,& \text{if } s_i = 1\\
    r_iH,              & \text{otherwise}
\end{cases}
\end{equation}
Since the commitments are homomorphically additive, the Pedersen commitment for the total assets is
\begin{equation}
C_{Assets}  = \left(\sum_{i=1}^{n} b_i \right)G + \left(\sum_{i=1}^{n} r_i \right)H = \sum_{i=1}^{n} c_i
\label{eq:C_assets}
\end{equation}

\subsubsection{Verification of the proof and total assets:}
The customer $\mathcal{C}$ (Verifier) of $\mathcal{E}$ verifies the proof $\pi_i$, for $i = 1, \dots, n$ along with the proof for $C_{Assets}$. $\mathcal{C}$ takes the public inputs $z_i = (y_i, v_i)$ which are the assignments to $\{a_{ij}\}_{j=0}^l$ and the proof $\pi = (A, B, C)$ to check \eqref{eq:verify}. The customer also needs to check commitment to $\mathcal{E}_{Assets}$ using $c_i$'s as per \eqref{eq:C_Assets}.

In Proof of Assets protocol, the exchange $\mathcal{E}$ convinces the verifier that
\begin{itemize}
\item If exchange knows the private key $x_i$, then $s_i = 1$ and the corresponding balance $v_i$ added to the commitment $C_{Asset}$.
\item The commitment $C_{Assets}$ is the commitment to the total assets $\sum_{i=1}^{n} s_iv_i$ of the exchange.
\end{itemize}

\subsection{Security and Privacy analysis}
In this section, we discuss the security properties of the proposed Proof of Assets protocol - completeness, sondness and statistical zero-knowledge. We also discuss the privacy of the exchange.

Consider a proof of assets protocol run between an exchange $\mathcal{E}$ and a customer $\mathcal{C}$. Let $out_{\mathcal{C}}^{PoA} \in \{Accept, Reject\}$ be the output of $\mathcal{C}$ based on the verification checks (a), (b) and (c) in verification phase of the protocol. 

\begin{lemma}
The above construction of the proof of assets protocol satisfies the completeness, sondness and statistical zero-knwoledge properties.
\end{lemma}
\proof
The proof for the security properties of the Proof of Assets protocol is straightforward from the construction of the ZK-SNARK proof system. \\ 
\textbf{Completeness:} The customer $\mathcal{C}$ needs to check \eqref{eq:verify_c_i} and \eqref{eq:verify_C_Assets} based on the verification step of the ZK-SNARK construction given in \eqref{eq:verify}. If the prover and verifier use the valid assignments for the coefficients $\{a_{ij}\}_{j=0}^m$ in computing \eqref{eq:pi} and \eqref{eq:verify}, i.e $(z_i, w_i) \in \mathcal{R}_{PoA}$, then the $b_{1,i} = 1$ and $b_{2,i} = 1$. Since $b_{2,i} = 1$, then \eqref{eq:C_Assets} also verified. It implies that $Pr[out_{\mathcal{C}}^{PoA} = Accept] = 1.$ \\
\textbf{Soundness:} Suppose an adversial exchange $\mathcal{A_{\mathcal{E}}}$ wants to prove invalid assignments to the witness $\{a_{ij}\}_{j=l+1}^{m} = w_i$ for a given $z_i$. For example, if $\mathcal{A_{\mathcal{E}}}$ change the values of the assignments corresponding to $s_i$ or the balance $v_i$ on his favour, then $(z_i,w_i) \not \in \mathcal{R}_{PoA}$ or $z_i \not \in L_{PoA}$.  The verification test fails in verifying the proof $\pi_i$ (i.e., $b_{1,i} = b_{2,i} = 0$). It implies that 
$Pr[out_{\mathcal{C}}^{PoA} = Accept] = 0.$ \\
\textbf{Statistical Zero-knowledge:} We need to show that there exist a polynomial time simulator $S$ which generates a proof (without knowing witness $w_i$) which is statistically identical to the real proof generated by $\mathcal{E}$ using $z_i$ and $w_i$.

The trapdoor information $\tau$ generated in the setup phase of the ZK-SNARK construction consists of elements that are drwan uniformly from the field $\mathbb{F}_p$. The field elements $U_i$, $V_i$ and $X_{wit,i}$ encoded in the proof $\pi_i$ as per \eqref{eq:pi} are statistically uniform. Where,
\begin{align*}
U_i = \sum_{j=0}^{m} a_{ij} u_j(x) , V_i = \sum_{j=0}^{m} a_{ij} v_j(x), \\  
X_{wit,i} =\sum_{j=l+1}^{m} a_{ij}\left(\beta u_j(x) + \alpha v_j(x) + w_j(x)\right)
\end{align*}

The prover uses the tuple $(z_i, w_i)$ in computing $U_i$, $V_i$ and $X_{wit,i}$. So, the field elements $A_i$, $B_i$ and $C_i$ of proof $\pi_i$ are from a uniform distribution. 
 
Consider the simulator $S$ in Table \ref{tab:S} for generating the proof $\pi_i$. The simulator picks random $A_i$ and $B_i$ from uniform distribution and determine the value of $C_i$ from $A_i$, $B_i$, $\sigma$ and $X_{pub,i}$ which is also distributed uniformly. Where, $X_{wit,i} =\sum_{j=0}^{l} a_{ij}\left(\beta u_j(x) + \alpha v_j(x) + w_j(x)\right)$, which is computed from $z_i = \{ a_{ij}\}_{j=0}^l$.
So, the simulated proof has also similar probabilty distribution to the real proof generated by $\mathcal{E}$ using the witness $w_i$.  Thus, the Proof of Assets protocol satisfies the statistical zero-knowledge.
\begin{table}[!t]
\caption{Simulator for generating proof $\pi_i$}
\begin{tcolorbox}[colback=white!4!white,colframe=white!25!black] 
\label{tab:S}
\begin{enumerate}
\item Picks $\alpha$, $\beta$, $\gamma$, $\delta$, $x \leftarrow \mathbb{F^{*}}$, set $\mathbf{\tau} = (\alpha$, $\beta$, $\gamma$, $\delta$, $x$)

\item Picks random polynomials for $Q$ and generates the CRS $\sigma = (\sigma_1, \sigma_2)$.

\item Pick $A_i$, $B_i$ $\leftarrow \mathbb{F}$
\item Determine $C_i$ from verification equation \eqref{eq:verify} using $A_i$, $B_i$, $\sigma$ and $\{a_{ij}\}_{j=0}^{l} = z_i$. 
\end{enumerate}
\end{tcolorbox}
\label{Simulator}
\end{table}

The privacy of $\mathcal{E}$ depends on the value of $s_i \in \{0,1\}$ as it captures if  $\mathcal{E}$ knows a private key or not. We define an experiment and we call  $AddressPrivacy^{\mathcal{D}}_{y_i,c_i}$. It denotes that for all $y_i \in \mathbf{S_{anon}}$ a distinguisher  $\mathcal{D}$ tries distinguish $y_i \in \mathbf{S_{own}}$ or not balance $v_i \in z_i$ and output commitment $c_i$. The experiment for a single instance $z_i \in L_{PoA}$ is defined as follows:
\begin{enumerate}
\item The public parameters are generated as $\mathbb{E},p, q, G, H \leftarrow Gen(1^{\lambda})$, where $\lambda$ is the security parameter.
\item $\mathcal{E}$ picks $b_i \leftarrow \{0,1\}$. 
\item $\mathcal{E}$ picks $r_i \in \mathbb{Z}_q$ and computes $c_i = b_iv_iG + r_iH$ and it determine the corresponding ZK-SNARK proof $\pi_i$ for $(z_i,w_i) \in \mathcal{R}_{PoA}$.
\item Let $z_i$, $c_i$ and $\pi_i$  are inputs to a polynomial time distinguisher $\mathcal{D}$. 
\begin{equation}
b'_i = \mathcal{D}(z_i,c_i,\pi_i)
\end{equation}
\item If $b'_i=b_i$, then $\mathcal{D}$ succeeds. Otherwise it fails. 
\end{enumerate}

\begin{definition}
For every $i \in [1,n]$ and $y_i \in \mathbf{S_{anon}}$, the proposed proof of assets protocol provides privacy of bitcoin address $y_i$ if for every PPT distinguisher $\mathcal{D}$ in the $AddressPrivacy^{\mathcal{D}}_{y_i,c_i}$ experiment with $Pr[AddressPrivacy^{\mathcal{D}}_{y_i,c_i} = 1] = Pr[b'_i = b_i] \leq \frac{1}{2} + negl(\lambda)$.
\end{definition}
The PPT distinguisher $\mathcal{D}$ can always toss a coin to guess $b'_i \in \{0,1\}$ (that gives a probability of $\frac{1}{2}$) or $\mathcal{D}$ can get a unique $r_i \in \mathbb{Z_q}$  to guess $b'_i$ with negligible probability $\frac{1}{|q|}$ (Where $|q|$ is polynomial in security parameter $\lambda$).  The above definition illustrates that for every $\mathcal{D}$ with inputs $z_i = (y_i, v_i)$ should not determine if the $y_i$'s balance $v_i$ is included in the $C_{Assets}$ with an advantage not more than negligibly close to $\frac{1}{2}$.  

\begin{table}[!t]
\caption{Performance of the single $C_{PoA}$ circuit}
\begin{center}
\begin{tabular}{|l|c|}
\hline
 Proof construction time & 0.2433 sec \\
 \hline
 Verififcation time &  0.0042 sec \\
 \hline
 Proof size  & $192$ bytes \\
\hline
\end{tabular}
\label{tab:tab3}
\end{center}
\end{table}

\begin{table*}[!t]
\caption{Performance of Proof of Assets protocol}
\begin{center}
\begin{tabular}{ |p{0.8cm}|p{1.8cm}|p{3.3 cm}|p{3.3cm}|p{2cm}|  }
 \hline
n & $|\mathbf{S_{own}}|$ ($\%$ $n$) & Construction time (seconds) & Verification time (seconds) & Proof size (MB) \\
\hline
$100$ & 25 & $16.96$ & $0.456$ & $0.01914$ \\
\hline
$100$ & 50 & $19.55$ & $0.455$ & $0.01914$ \\
\hline
$100$ & 75 & $24.76$ & $0.455$ & $0.01914$ \\
\hline
$1000$ & 25 &$166.026$ & $4.528$ & $0.1914$ \\
\hline
$1000$ & 50 &$191.942$ & $4.514$ & $0.1914$ \\
\hline
$1000$ & 75 &$217.934$ & $4.512$ & $0.1914$ \\
\hline
$10000$ & 25 &$1657.54$ & $45.36$ & $1.914$ \\
\hline
$10000$ & 50 &$1913.79$ & $45.07$ & $1.914$ \\
\hline
$10000$ & 75 &$2196.55$ & $45.09$ & $1.914$ \\
\hline
\end{tabular}
\label{tab:performance}
\end{center}
\end{table*}

\begin{lemma}
The proposed proof of assets protocol provides privacy of addresses owned by $\mathcal{E}$ under the discrete log (DL)~\cite{DL} assumption.
\end{lemma}
\proof
Suppose an adversarial exchange $\mathcal{A_{\mathcal{E}}}$ wants to solve a DL problem. $\mathcal{A_{\mathcal{E}}}$ picks $b_i \in \{0,1\}$ and determine $c_i = b_iv_iG + r_iG$ (step 2 in $AddressPrivacy^{\mathcal{D}}_{y_i,c_i}$ experiment). It gives $(z_i= (y_i, v_i), c_i and \pi_i)$ as input to  $\mathcal{D}$. Since, $\mathcal{D}$ is a PPT algorithm, $\mathcal{A_{\mathcal{E}}}$ also a PPT algorithm. \\
The generators $G$ and $H$ are chosen uniformly and independetly from $\mathbb{E}$. We have $G = kH$, for some unknown $k$ ($k$ is not known to $\mathcal{A_{\mathcal{E}}}$).
\begin{equation}
y_i = x_iG, \hspace{0.2cm}  c_i= 
\begin{cases}
    (kv_i+r_i)H,& \text{if } b_i = 1\\
    r_iH,              & \text{otherwise}
\end{cases}
\end{equation}
If $b'_i = \mathcal{D}(z_i,c_i, \pi_i)$, then $\mathcal{A_{\mathcal{E}}}$ outputs $b'_i$.
Suppose $\mathcal{D}$'s success probability $Pr[b'_i = b_i] > \frac{1}{2} + negl(\lambda)$, then $\mathcal{A_{\mathcal{E}}}$)'s success probability is also larger than   $\frac{1}{2} + negl(\lambda)$. The PPT adversary can solve the DL problem, this is a contradiction. Thus, $\mathcal{D}$ has an advantage less than $\frac{1}{2} + negl(\lambda)$ to reveal the privacy of the exchange.
\section{Implementation and Performance evaluation}
We have implemented the Proof of Assets protocol in $C++$ using the \textit{libsnark} library \cite{libsnark} developed by scipr-lab. 
The implementation consists of the following elements.\\
\textbf{Protoboard.} A protoboard is a virtual prototype which collect all the circuits similar to a prototyping board to attch all the circuits and chips in the electronic circuit board. We need to allocate  all the public and auxiliary inputs used in the ZK-SNARK proof system to the protoboard. Protoboard is defined as 
\begin{align*}
protoboard<FieldT> pb;
\end{align*}
\textbf{Gadgets.} The \textit{libsnark} library provides several gadgets. For example, we use \textit{packing gadget} (for unpacking a scalar $x_i$ to $xvec_i$ as discussed in section \ref{sec:C_MUL}) in this work. We construct a \textit{PoA\_gadget} to implement circuit $C_{PoA}$ for \textit{PoA}. The \textit{PoA\_gadget} gadget checks the correctsness of the NP-staement \textit{PoA} and is defined as 
\begin{align*}
PoA\_gadget<FieldT> POA(pb, public \hspace{0.1cm} inputs, witness);
\end{align*}
and generate the R1CSS constraints,
\begin{align*}
POA \rightarrow generate\_r1cs\_constarints();
\end{align*}
Finally, after giving all the input (public and witness) values by the prover ($\mathcal{E}$) the following function generates the witness to all the internal wires of the circuit $C_{PoA}$ and outputs Pedersen commitment $c_i$.
\begin{align*}
POA \rightarrow generate\_r1cs\_witness();
\end{align*}

We divide the implementation of \textit{PoA\_gadget} for \textit{PoA}  into three subcircuits or gadgets. The scalar multiplication gadget (\textit{scalr\_mul}) verifies the knowledge of $\mathcal{E}$ on private key $x_i$. The comparison gadget (\textit{cmp\_gadget}) checks the equality of the computed public key ($y'_i$) from the given input address ($y_i$). Finally, the  Pedersen commitment gadget (\textit{Pedersen\_gadget}) verifies the commitment $c_i$ to the balance $v_i$. Each of these gadgets consists for functions for generating $R1CS$ constriants and generating witness similar to the main gadget. The \textit{PoA\_gadget} generates a total of $1702$ constraints for each instance $(z_i,w_i) \in \mathcal{R}_{PoA}$.

 We performed tests on a personal computer with Intel(R) Core(TM) $i9-9900K$ CPU $@$ $3.60$GHz processor with $16$GB RAM using a single core.
The details of the proof construction time by an exchange $\mathcal{E}$, proof verification time by the customer $\mathcal{C}$, and the proof size are described in Table \ref{tab:tab3}. The proof construction time depends on the number of the ZK-SNARK circuit's constraints. The proof size is the combination of size of $3$ group elements of each proof $\pi_i$ and size of the Pedersen commitment to $v_i$.

Table \ref{tab:performance} illustrates the performance of the Proof of Assets protocol with the size of the set $\mathbf{S_{anon}}$ ($n$).  We test the protocol for $n = 100, 1000$ and, $10000$. We choose $|\mathbf{S_{own}}|$ as $25\%$, $50\%$ and $75\%$ of $n$. We assume $\mathcal{E}$ provides proofs for $n-|\mathbf{S_{own}}|$ number of addresses by considering $s_i = 0$ for all $y_i \not \in \mathbf{S}_{own}$ to reduce the proof construction time.
The proof construction time includes the time required for the construction of the proofs for $n$ number of $C_{PoA}$  and the time required to generate the proof for the Pedersen commitment $C_{Assets}$ \eqref{eq:C_assets}. Similarly, for the proof verification and proof size. 

The construction time, verification time, and proof size increases linearly with $n$. The results show that the Proof of Assets protocol is efficient in practice as the regular PC constructs the proof in less than an hour and the proof size is less than $\approx 15$ MB for $n = 10000$ with short verification time. The performance of the protocol will be improved on servers with high-end processors.

\section{Conclusions and Future Research}
In this paper, we described the ZK-SNARK based proof
of assets protocol for bitcoin exchanges by preserving the
privacy of the exchanges without revealing the public keys or the balances associated with the public keys. This is achieved by proving the knowledge of the private keys associated with the public keys (Bitcoin P2PK addresses) using the ZK-SNARK mechanism with Pedersen commitment as the output of the circuit. We also analyse the security and privacy properties of the proposed protocol. Through the
simulation results, we showed the efficiency of the protocol for proof construction, verification and proof size. In the future, we foresee  the construction of the proof of assets protocol for bitcoin P2PKH (Pay to Public Key Hash) addresses by proving the knowledge of the hash preimage through the ZK-SNARK framework. We may also combine these proof of assets protocols with proof of liabilities by proving the membership of customer funds using the set-membership proofs and ZK-SNARK mechanism.

\bibliography{IEEEabrv,TSMC.bib}

\end{document}

%% file: figs/ZK-SNARK.tex
%

\begin{tikzpicture}[auto, node distance=2cm,>=latex']
\node [text width=10em, text centered, minimum height=1em,name=genesis] (A) {$L$: NP-statement};

\node [text width=5em, text centered, minimum height=3em, below of=A,node distance=2cm] (B) {circuit $C$};

\draw [double,->,thick] (A) -- node {} (B);

\node [text width=10em, text centered, minimum height=3em, below of=B,node distance=2cm] (C) {Arithmetic/Boolean Gates};

\draw [double,->,thick] (B) -- node {} (C);

\node [text width=10em, text centered, minimum height=3em, below of=C,node distance=2cm] (D) {$R1CS$};

\draw [double,->,thick] (C) -- node {} (D);

\node [text width=5em, text centered, minimum height=3em, below of=D,node distance=2cm] (E) {QAP/QSP};

\draw [double,->,thick] (D) -- node {} (E);

\node [text width=8em, text centered, minimum height=3em, below of=E,node distance=2cm] (F) {ZK-SNARK};

\draw [double,->,thick] (E) -- node {} (F);
\end{tikzpicture}

%% file: figs/NP.tex
%

\begin{tikzpicture}[auto, node distance=2cm,>=latex']
\node [text width=10em, text centered, minimum height=1em,name=genesis] (A) {\textbf{$x_i$}};

\node [text width=5em, text centered, minimum height=1em, right of=A,node distance=2cm] (B) {$y_i$};

\node [rectangle, draw, text width=5.5em, text centered, minimum height=3em, below of=A,node distance=2cm] (C) {$C_{MUL}(x_i,G)$ };

\draw [->,thick] (A) -- node {} (C);

\node [rectangle, draw, text width=6em, text centered, minimum height=3em, below of=B,node distance=4cm] (D) {$C_{CMP}(y'_i,y_i)$};

\node [text width=-8em, text centered, minimum height=-8em, below of=C,node distance=2cm] (E) {};

\draw [->,thick] (C) |- node {$ \hspace{-2cm}\vspace{-1cm} y_i^{'} = x_i.G$} (D);
\draw [->,thick] (B) -- node {} (D);

\node [text width=8em, text centered, minimum height=1em, right of=B,node distance=3cm] (F) {$v_i = bal(y_i)$};

\node [rectangle, draw, text width=4em, text centered, minimum height=3em, below of=F,node distance=4cm] (G) {$s_i*v_i$};

\draw [->,thick] (D) -- node {\textbf{$s_i$}} (G);

\node [rectangle, draw, text width=7em, text centered, minimum height=3em, below of=G,node distance=2.5cm] (H) {$C_{PED}(b_i,r_i)$};

\draw [->,thick] (F) -- node {} (G);
\draw [->,thick] (G) -- node {$b_i$} (H);

\node [text width=0em, text centered, minimum height=0em, below right of=G,node distance=2.5cm] (I) {$r_i$};
\draw [->,thick] (I) |- node {} (H);
\node [text width=7em, text centered, minimum height=0em, below of=H,node distance=2cm] (I) {$c_i = b_iG + r_iH$};
\draw [->,thick] (H) -- node {} (I);


\end{tikzpicture}